\documentclass[twocolumn,prb,amsmath,amssymb]{revtex4}

\usepackage[dvipdfm]{graphicx}
\usepackage{dcolumn}
\usepackage{bm}
\usepackage{amsmath}

\begin{document}

\preprint{APS/123-QED}

\title{Terahertz conductivity spectroscopy of Co-doped BaFe$_2$As$_2$ Thin Film}

\author{D. Nakamura}
\author{Y. Imai}%
\author{A. Maeda}%
\author{T. Katase$^1$}%
\author{H. Hiramatsu$^2$\footnote{Present address: Frontier Research Center, Tokyo Institute of Technology, Japan}}
\author{H. Hosono$^{1,2,3}$}%
\affiliation{%
Department of Basic Science, the University of Tokyo, 3-8-1, Komaba, Meguro-ku, Tokyo, 153-8902, Japan\\
JST, TRIP, Sanbancho, Chiyoda-ku, Tokyo 102-0075, Japan\\
$^1$Materials and Structures Laboratory, Tokyo Institute of Technology,
Mail-box R3-1, 4259 Nagatsuta-cho, Midori-ku, Yokohama 226-8503, Japan\\
$^2$ERATO-SORST, Japan Science and Technology Agency (JST), in Frontier Research Center, Tokyo Institute of Technology,
S2-6F East, Mail-box S2-13, 4259 Nagatsuta-cho, Midori-ku, Yokohama 226-8503, Japan\\
$^{3}$Frontier Research Center, Tokyo Institute of Technology,
S2-6F East, Mail-box S2-13, 4259 Nagatsuta-cho, Midori-ku, Yokohama 226-8503, Japan}%

\date{22 December 2009}

\begin{abstract}
We investigated the complex conductivity spectrum of a Co-doped BaFe$_2$As$_2$ epitaxial thin film in the THz region.
In the normal state, the complex conductivity shows a Drude-type frequency dependence, while in the superconducting state, the frequency dependence of the complex conductivity changes to that of a typical superconducting materials.
We estimated the magnetic penetration depth at absolute zero to be 710 nm and the superconducting gap energy to be 2.8 meV, which is considered to be the superconducting gap opened at the electron-type Fermi surface near the M point.
We succeeded in obtaining the low-energy elementary excitation of a Fe-based superconductor using the electromagnetic method without invoking the Kramers-Kronig transformation.
\end{abstract}

\maketitle

Fe-based superconductors have stimulated significant interest and competition for the synthesis of a novel type of superconductor since the discovery of superconductivity in F-doped LaFeAsO in 2008 \cite{Kamihara08}.
Currently, Fe-based superconductors are classified into five families based on the structure.
Among them, $R$FeAsO ($R$ : rare earths) and $Ae$Fe$_2$As$_2$ ($Ae$ = Ca, Sr, Ba, Eu, K) systems have been widely investigated \cite{Ishida09}.
The highest superconducting transition temperature, $T_c$ $\sim$ 55 K, is found in F-doped $R$FeAsO \cite{Ren08}.
Yet, the $Ae$Fe$_2$As$_2$ system is superior for thin film application, due to its relatively high $T_c$ and the existence of only one kind of anion, which makes the fabrication of the high-quality thin film easier than in the $R$FeAsO system.
\vspace{-0.1cm}

As for the $Ae$Fe$_2$As$_2$ system, the undoped BaFe$_2$As$_2$ possesses metallic conductivity \cite{Rotter08}, and Co-doping of the Fe-site leads to electron-doped superconductivity at 22 K \cite{Sefat08}.
With decreasing temperature, structural and antiferromagnetic phase transitions occur at 134-140 K\cite{Chu09, Rotter08v2} that are accompanied by an anomaly in the DC resistivity.
These phase transitions are observed by the DC resistivity, up to 6 \% Co-doping in the electronic phase diagram\cite{Chu09, Wang09, Ning09}.
Superconductivity occurs at 2.5 - 18 \% Co-doping, which indicates that there is a subtle relationship between the antiferromagnetic phase and the superconducting phase in the underdoped BaFe$_2$As$_2$.
The Fermi surface of Fe-based superconductors is much more complicated than that of high-$T_c$ cuprate superconductors because of their multiband nature.
A band calculation study clarified that there are hole-type and electron-type Fermi surfaces around the $\Gamma$ and M points in a reciprocal space, respectively \cite{Kemper09}.
The superconducting gap opens at both Fermi surfaces in the superconducting state, which causes the novel symmetry of the superconducting gap, $s\pm $ \cite{Mazin08}.
By the thermal conductivity \cite{Tanatar09}, the microwave conductivity \cite{Hashimoto09}, and the ARPES \cite{Terashima09}, the existence of the nodeless gap in the superconducting state is suggested, which is consistent with the $s\pm $ symmetry.
The Fermi-surface-dependent superconducting gap energies, which are identified in a Ba(Fe$_{0.925}$Co$_{0.075}$)$_2$As$_2$ single crystal ($T_c = 25.5 $ K) by ARPES \cite{Terashima09}, are 6.7 meV at the "$\Gamma$" surface and 4.5 meV at the "M" surface, respectively.
\vspace{-0.1cm}

For the fabrication of $Ae$Fe$_2$As$_2$ thin films, we succeeded in obtaining Co-doped SrFe$_2$As$_2$ epitaxial thin films \cite{Hiramatsu08}.
Because the films are quite sensitive to water vapor in ambient air \cite{Hiramatsu09}, it is not easy to measure their superconducting properties.
However, we recently found that Co-doped BaFe$_2$As$_2$ thin films are much more stable in air than the Sr-based thin films \cite{Katase09}, which increases the possibility of studying their physical properties, including the superconducting gap energy, $\Delta_{\text{sc}}$.
The energy scale of $\Delta_{\text{sc}}$ is close to the terahertz (THz) frequency range, thus we investigated the THz conductivity of Co-doped BaFe$_2$As$_2$ thin film in this paper.


\begin{figure}[t]
\begin{center}
\begin{minipage}{0.64\linewidth}
\includegraphics[width=0.99\linewidth]{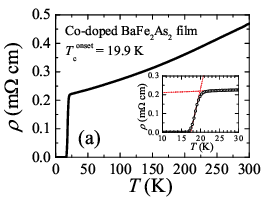}
\end{minipage}
\begin{minipage}{0.34\linewidth}
\includegraphics[width=0.99\linewidth]{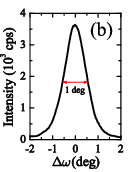}
\end{minipage}
\vspace{-0.7cm}
\includegraphics[width=0.9\linewidth]{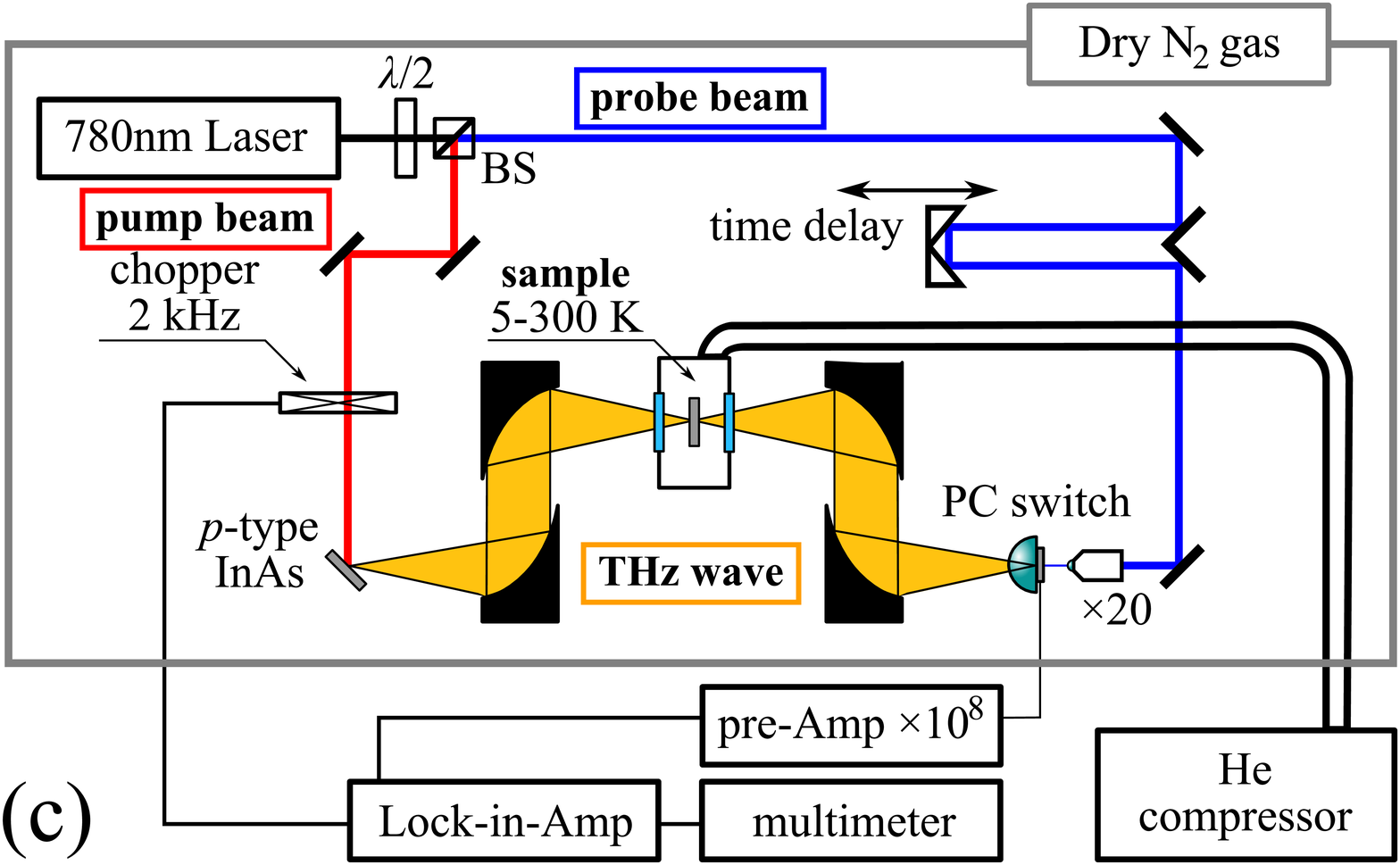}\\
\vspace{0.5cm}
\caption{\label{label} (color online). 
(a) The temperature dependence for the DC resistivity of Co-doped BaFe$_2$As$_2$ thin film. The inset figure shows the enlarged plot of the region near $T_c$. The onset temperature of the superconducting transition, $T_c^{\text{onset}}$, was determined from the crosspoint of the dotted additional lines.
(b) The rocking curve of the (004) reflection. 
(c) The schematic figure of the transmitted-type THz time-domain spectroscopy. Details are described in the text.
}
\vspace{-0.5cm}
\end{center}
\end{figure}

Co-doped BaFe$_2$As$_2$ epitaxial thin film was grown on a MgO(001) substrate at 730$\ {}^\circ\mathrm{C}$ by a pulsed laser deposition technique using a BaFe$_{1.8}$Co$_{0.2}$As$_2$ disk target \cite{Katase09}.
The overall dimension of the film was 5$\times$10 mm$^2$, and the film thickness was 110 nm, which was measured using a surface profiler.
Figure 1(a) shows the temperature dependence of the DC resistivity.
There is no anomaly indicating the antiferromagnetic phase.
From the enlarged plot near $T_c$ (the inset of Fig. 1(a)), the onset temperature of the superconducting transition, $T_c^{\text{onset}}$, is estimated as 19.9 K.
The midpoint of the transition is 18.7 K, and the transition width is approximately 2.9 K. 
The X-ray diffraction measurements indicate that the epitaxial relationship between the film and the substrate is BaFe$_2$As$_2$(001) $\parallel $ MgO(001) for out-of-plane, and BaFe$_2$As$_2$[100] $\parallel $ MgO[100] for in-plane.
Figure 1(b) is the rocking curve for the (004) reflection of the film, while the full width at half maximum is about 1 deg.

The frequency-dependent (0.2-1.5 THz) complex conductivity was measured by the transmitted-type THz time-domain spectroscopy (THz-TDS) \cite{Grischowsky}, using the measurement system shown in Fig. 1(c).
For the emitter and the detector of the THz-wave, we used a Zn-doped InAs crystal and a dipole-type photoconductive switch, respectively.
By measuring the photocurrent of the THz detector as a function of the optical path length of the probe beam (blue line in Fig. 1(c)), we can obtain the time-domain waveform of the THz pulse.
Two time-domain waveforms of the transmitted THz-wave through the sample (Co-doped BaFe$_2$As$_2$ film on a MgO substrate) and the reference (MgO substrate only) were measured.
In the cryostat containing quartz optical windows, we set the sample holder such that the direction of the electric field was parallel to the FeAs plane.
The working surface area of the film was approximately 1$\times$1 mm$^2$, where the THz pulse was irradiated.
The measurement system was continuously purged with dry nitrogen to eliminate the effect of the absorption spectrum of water.

We obtained the frequency-dependent transmittance, $T(\omega)$ ($\omega$ is the angular frequency), and the phase shift, $\Delta\phi (\omega)$, by dividing the Fourier-transformed signal of the sample by that of the reference.
The complex refractive index of the film, $\tilde n \equiv n - i\kappa$, was obtained by fitting the data to the analytical formula \cite{Hangyo02}.
\vspace{-0.5cm}

\begin{equation}
\begin{split}
 \sqrt{T(\omega)} \exp [-i\Delta \phi (\omega)]  &= \\
 \frac{2\tilde n (\tilde n_{\text{sub}}+1)}{(\tilde n + 1)(\tilde n + \tilde n_{\text{sub}})} & \frac{\exp \left[-i \frac{(\tilde n -1) d\omega}{c} \right]}{1-\frac{\tilde n - 1}{\tilde n + 1}\frac{\tilde n - \tilde n_{\text{sub}}}{\tilde n + \tilde n_{\text{sub}}} \exp \left[-i \frac{2\tilde n d\omega}{c} \right] } ,
\end{split}
\end{equation}

where $\tilde n_{\text{sub}}$ is the complex refractive index of the MgO substrate, $c$ is the speed of light, and $d$ is the thickness of the film, respectively.
Then, the complex dielectric function, $\tilde \varepsilon \equiv \varepsilon_1 - i\varepsilon_2 = \varepsilon_0 \tilde n^2$, where $\varepsilon_0$ is the permittivity of a vacuum, and the complex conductivity, $\tilde \sigma \equiv \sigma_1 + i\sigma_2 = i \omega (\tilde \varepsilon - \varepsilon_0)$, were calculated.
We separately measured $\tilde n_{\text{sub}}$ for the MgO substrate as $n_{\text{sub}} \sim 3.18$ and $\kappa_{\text{sub}} \sim 0.006$, which are consistent with the previously reported values \cite{Murakami02}.
Therefore, $\tilde n (\omega)$ can be obtained without using the Kramers-Kronig transformation.
For the THz-TDS method, the reliable amplitude and the phase can be obtained in a single experiment, without performing numerrous experiments over a wide frequency interval as an usual optical measurement.

\begin{figure*}[ht]
\begin{center}
\begin{minipage}{0.32\linewidth}
\includegraphics[width=0.99\linewidth]{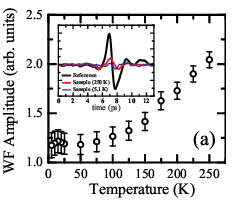}
\end{minipage}
\hspace{0.3cm}
\begin{minipage}{0.32\linewidth}
\includegraphics[width=0.99\linewidth]{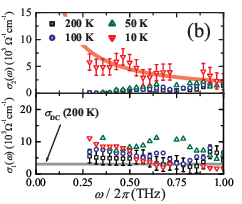}
\end{minipage}
\begin{minipage}{0.32\linewidth}
\includegraphics[width=0.99\linewidth]{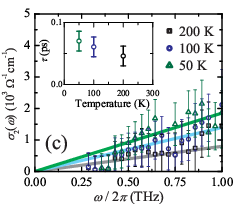}
\end{minipage}
\caption{\label{label} (color online). 
(a) The temperature dependence of the amplitude of the transmitted THz waveform through the sample. The inset shows the time-domain waveforms of the THz pulse through the reference (thickest solid line) and through the sample (second thickest line: 250 K, thinnest line: 5.1 K).
(b) The frequency dependence of the complex conductivity. The lower and the upper panel represent the real and the imaginary parts, respectively. The solid line in the bottom panel represents the DC conductivity (3$\times 10^3 \ \Omega^{-1} $cm$^{-1}$) at 200 K, while the solid line in the upper panel indicates the $\omega^{-1}$ fitting curve.
(c) An enlarged plot of $\sigma_2 (\omega)$ at $T > T_c$. The solid lines are linear fitting curves. The inset is the temperature dependence of the lifetime of the carrier, which is estimated from the slope of $\sigma_2(\omega)$.
\vspace{-0.5cm}}
\end{center}
\end{figure*}

In the transmitted-type THz-TDS, the difference in the thickness of the substrate, $\delta d$, causes a serious error in $\Delta \phi$.
To eliminate the error in the phase shift, we corrected the obtained phase shift in the following manner.
Because $n_{\text{sub}} \gg \kappa_{\text{sub}}$, $\delta d$ has almost no influence on the transmittance.
If we assume $\delta d$ = 20 $\mu $m, the error in the transmittance is roughly estimated by $1- \exp [-2\omega \kappa_{\text{sub}} \delta d / c] = 0.005$ at 1 THz.
Therefore, the error from the experimentally obtained transmittance can be ignored, and only the experimentally obtained phase shift, $\Delta \phi_{\text{exp}}$, requires calibration.
In addition, we assumed that $n = \kappa $ at the highest temperature (250 K).
This indicates that $\sigma_2 = \omega \varepsilon_0 (n^2 -\kappa^2) = 0$.
This corresponds to the low-frequency response of the Drude metal ($\omega < \tau^{-1} \ll \omega_p$) and is widely used for the discussion of the microwave conductivity of metallic materials \cite{Bonn1996}.
Therefore, from the experimentally obtained $T(\omega)$ and $n-\kappa = 0$, we calculated $(n+\kappa, \Delta\phi_{n=\kappa})$ from eq. (1) at each frequency.
Then, we extracted $\delta d$ as $( \Delta \phi_{\text{exp}} - \Delta\phi_{n=\kappa} ) c / \omega n_{\text{sub}}$, and subtracted $\Delta \phi_{\text{exp}} - \Delta\phi_{n=\kappa}$ from $\Delta \phi_{\text{exp}}$ over the entire temperature region.
$\delta d$ was then determined to be 23.51 $\pm$ 6.6 $\mu$m, which is consistent with the value roughly measured by the micrometer.
Therefore, we believe that our treatment is appropriate.
After this calibration, we calculated the $\tilde \sigma (\omega)$ of the film.


Figure 2(a) shows the temperature dependence of the amplitude of transmitted THz pulses through the Co-doped BaFe$_2$As$_2$ film at $T$ = 5.1-250 K.
The inset of Fig. 2(a) shows the time-domain waveforms of the THz pulse through the reference and the sample.
In the normal state, the amplitude of the transmitted THz pulse decreases with temperature, which is based on the reduction of the DC resistivity.
In the superconducting state, the amplitude of the transmitted THz pulse is suppressed to a great extent.
These types of temperature dependence are found in many known superconductors \cite{Murakami02}.
Figure 2(b) shows $\tilde \sigma (\omega)$ (lower panel is the real part, while upper panel is the imaginary part).
Under conditions where $T > T_c$, $\sigma_1 \sim \sigma_{\text{DC}}$ ($\sigma_{\text{DC}}$ is the DC conductivity, which is represented by the solid line (at $T$ = 200 K) in the bottom panel of Fig. 2(b)), and is almost independent of frequency.
In addition, $\sigma_2$ is almost equivalent to zero and slightly increases with frequency.
Therefore, $\tilde \sigma (\omega)$ resembles that of Drude-type metal.
In Fe-based superconductors, there are two kinds of carriers (holes at the Fermi surface near the $\Gamma$ point and electrons at the Fermi surface near the M point).
Therefore, the experimentally obtained conductivity is the summation of the conductivity of both the electrons and the holes.
However, for the electron-doped system, the volume of the hole pocket decreases with increasing doping level, in contrast to that of the electron pocket.
For BaFe$_{1.8}$Co$_{0.2}$As$_2$, the number of the electrons is approximately twice as large as that of the holes \cite{Fang09}.
Therefore, we estimate that the life-time of the electron is reflected in the measured value $\sigma_2 (\omega)$, as $\tau = \sigma_2 (\omega) / \sigma_{\text{DC}} \omega$, which is shown in the inset of Fig. 2(c).
The obtained plasma frequency, $\omega_p = \sqrt{\sigma_{\text{DC}} / \epsilon_0 \tau}$, is about 0.56 eV at $T$ = 200 K, and this frequency does not almost depend on the temperature from 50 K to 200 K.

As the temperature decreases to a value below $T_c$, the $\tilde \sigma (\omega)$ factor of a typical superconducting material can be observed.
$\sigma_1$ is suppressed over the entire frequency region and $\sigma_2$ is significantly enhanced compared with that in the normal state.
We note that there remains some finite conductivity ($\sim 4\times 10^3 \ \Omega^{-1}$cm$^{-1}$) in the superconducting state, which is qualitatively similar to the result of an optical measurement \cite{Gorshunov09}.
We have some possibilities for the origin of the finite conductivity in the superconducting state.
(1) the residual conductivity by uncondensed carriers not because of the presence of nodes but because of the dirty nature of the film we used, (2) the residual conductivity owing to the presence of nodes or anisotropy, $etc.$
While we can not find the decisive origin, the following discussion about the penetration depth suggests that the former possibility might be more likely in our film.
$\sigma_2 (\omega)$ can be represented as $\sigma_2 (\omega) \sim 1 / \mu_0 \omega \lambda^2 (T)$ (the thick line in the upper panel of Fig. 2(b) is the $\omega^{-1}$ fitting curve), and we estimated the magnetic penetration depth at absolute zero ($\lambda(0 \text{ K}) \sim$ 710 $\pm$ 70 nm) assuming the temperature dependence of the phenomenological penetration depth is $\lambda(0\text{ K}) = \lambda(T) \sqrt{1-(T/T_c)^4}$.
If we calculate the superconducting plasma frequency as $\omega_{ps} = c/\lambda$, $\omega_{ps}$ = 0.28 eV.
Because the carrier density is proportional to the square of the plasma frequency, this result suggests that 25\% of the normal carrier contributes to the superfluid density in our film. 
Therefore, we suggest that the former possibility about the finite conductivity in the superconducting state might be more likely, in the present stage.
The penetration depth of our film is larger than the previously reported value, $\lambda$(0 K) = 208 nm (from $H_{c1}$ \cite{Gordon2009}), $\lambda$(1.7 K) = 245 nm (from muon spin rotation \cite{Williams09}), $\lambda$(10 K) = 350 nm and $\lambda$(0 K) = 360 nm (from optical measurement \cite{Heumen09, Gorshunov09}).
While a slightly broad superconducting transition ($\Delta T_c \sim$ 2.9 K) might be also related to a larger $\lambda$, there is a report for the classical superconductor, lead, that $\lambda$ of films is approximately 10 times larger than that of bulk sample \cite{Glover57}.
Therefore, we think that the interpretation for this issue deserve further studies, using various different films.
We note that this will not affect the essential aspect of the following discussions regarding the superconducting gap.

\begin{figure*}[htbp]
\begin{center}
\begin{minipage}{0.29\linewidth}
\includegraphics[width=0.99\linewidth]{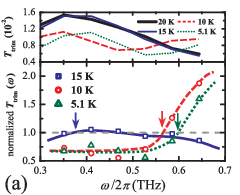}\\
\end{minipage}
\hspace{-0.3cm}
\begin{minipage}{0.27\linewidth}
\includegraphics[width=0.99\linewidth]{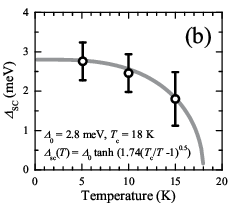}
\end{minipage}
\begin{minipage}{0.42\linewidth}
\caption{\label{label} (color online). 
(a) (upper panel) The frequency dependence of the transmittance of the extracted time-domain waveform. (bottom panel) The normalized transmittance by the transmittance directly above $T_c$ (20 K). The lines are the eyeguide, while the down arrows indicate the characteristic frequency where the transmittance increases in comparison to that of 20 K.
(b) The temperature dependence of the superconducting gap energy, $\Delta_{\text{sc}}$, which is estimated from the characteristic frequency (down arrows in Fig. 3(a)). The solid line is the fitting curve of the BCS-type temperature dependence of $\Delta_{\text{sc}}$, $\Delta_{\text{sc}}(T) = \Delta_0 \tanh (1.74(T_c / T -1)^{0.5})$.
}
\end{minipage}
\vspace{-0.5cm}
\end{center}
\end{figure*}

Next, we estimated the superconducting gap energy, $\Delta_{\text{sc}}$.
At $T < T_c$, the transmitted signal through the film is weak and noisy, so we extracted the data in the vicinity of main peak (4 ps $\leq  t \leq$ 11 ps) from the time-domain waveform.
The upper panel of Fig. 3(a) shows the frequency dependence of the transmittance, $T_{\text{trim}} (\omega)$, which is Fourier-transformed from the extracted time-domain waveform.
By using this procedure, the noisy data in the original time-domain waveform (at $t$ $<$ 4 ps and $t$ $>$ 11 ps) does not affect the Fourier-transformed spectrum such that we can discuss the physical nature in the superconducting state more clearly.
Compared with $T_{\text{trim}} (\omega)$ just above $T_c^{\text{onset}}$ ($T$ = 20 K), $T_{\text{trim}} (\omega)$ below $T_c^{\text{onset}}$ was enhanced at some frequencies.
The bottom panel of Fig. 3(a) indicates the normalized transmittance, $T_{\text{normalized}} (\omega, T)  = T_{\text{trim}} (\omega, T) / T_{\text{trim}} (\omega, T$ = 20 K).
The solid lines are the eyeguide.
From this panel we can easily identify the characteristic frequency, $\omega^\ast$, where $T_{\text{normalized}} (\omega) = 1$ (down arrows).
Notably, $\omega^\ast$ gradually increases with decreasing temperature, suggesting that $\hbar \omega^\ast$ is related to the energy scale of the superconducting gap.
In dirty-limit BCS superconductors, Palmer and Tinkham calculated the frequency dependence of the transmission ratio \cite{Palmer68}.
According to their calculation, $\hbar \omega^\ast $ is close to $1.4 \Delta_{\text{sc}}$, which leads to $\Delta_{\text{sc}}(T = 0) = 2.0 $ meV in the case of our film.
For strong-coupling superconductors, however, $\hbar \omega^\ast \sim \Delta_{\text{sc}}$.
For the $Ae$Fe$_2$As$_2$ system, the strong coupling is suggested by a $\mu$SR study\cite{Hiraishi09}.
Therefore, we regard that $\hbar \omega^\ast \sim \Delta_{\text{sc}}$, and subsequently plot the temperature dependence of $\Delta_{\text{sc}}$ in Fig. 3(b).
The solid curve indicates the fit to the BCS-type function, $\Delta_{\text{sc}} (T) = \Delta_{\text{sc}}(0) \tanh (1.74 (T_c / T -1)^{0.5} )$, which is also applicable to strongly-coupled superconductors \cite{Palmer68}.
Using these fitting parameters ($\Delta_{\text{sc}} (0)$ =  2.8 $\pm$ 0.5 meV and $T_c$ = 18 K) we obtained $2\Delta_{\text{sc}} (0) / k_B T_c$ = 3.6 $\pm$ 0.6.
We presume that the difference between $T_c$ determined by this fit and that obtained in the DC resistivity measurement originates from the slight change of quality in the sample handling in ambient air.
According to the inset of Fig. 2(c), $h / \tau$ is about 59 meV at $T$ = 50 K.
Therefore, we can conclude that our sample is close to the dirty limit ($\Delta_{\text{sc}} (0) \ll  h / \tau$).
This result also supports the validity of the above estimation of $\Delta_{\text{sc}}$ for the dirty and strong-coupling material.
Using ARPES measurement, Terashima $et$ $al.$ \cite{Terashima09} reported that there are two superconducting gap energies in a BaFe$_{1.85}$Co$_{0.15}$As$_2$ single crystal ($T_c$ = 25.5 K).
These gaps include the small superconducting gap, $\Delta_S (0) \sim $ 4.5 meV, opens at the electron-type Fermi surface near the M point, and the large superconducting gap, $\Delta_L (0) \sim $ 6.7 meV, opens at the hole-type Fermi surface near the $\Gamma$ point.
The value of $2\Delta_S (0) / k_B T_c$ is close to the value we obtained, 3.6 $\pm$ 0.6 ($2\Delta_S (0) / k_B T_c$ = 4.1 and $2\Delta_L (0) / k_B T_c$ = 6.1).
Therefore, we concluded that the enhanced part of $T_{\text{trim}}(\omega)$ at $T < T_c$ originates from the superconducting gap opened at the Fermi surface near the M point.
The large superconducting gap opened at the Fermi surface near the $\Gamma$ point exceeds our experimental frequency range (6.7 meV corresponds to 1.63 THz).
In completion of this work, quite recently an FIR measurement with using the Kramers-Kronig transformation \cite{Heumen09} reported $\sigma_1 (\omega)$ indicating the existence of the superconducting gap at 3.1 meV and 7 meV.
Also, the optical measurement with using the Mach-Zehnder arrangement \cite{Gorshunov09} reports the single small superconducting gap (1.85 meV).
The superconducting gap we obtained is consistent with that of optical conductivity \cite{Heumen09}, and $\hbar \omega^\ast$ corresponds to the smaller superconducting gap. 

In conclusion, we measured the complex conductivity of Co-doped BaFe$_2$As$_2$ epitaxial thin film in the THz region, and we observed the Drude-type complex conductivity in the normal state.
In the superconducting state, $\tilde \sigma (\omega)$ of the film changes to the typical $\tilde \sigma (\omega)$ of the superconducting material.
We estimated the magnetic penetration depth at absolute zero as 710 nm.
Also, we estimated the superconducting gap energy as 2.8 meV, which is considered comparable to the energy gap opened at the Fermi surface near the M point.
We succeeded in obtaining the low-energy elementary excitation of Fe-based superconductor using the electromagnetic method without invoking Kramers-Kronig transformation.

We thank Prof. M. Tonouchi for the photoconductive switch.
We also thank the Japan Society for the Promotion of Science for the financial support of D. Nakamura.
H. Hiramatsu and H. Hosono are supported by the Funding Program for World-Leading Innovative R\&D on Science and Technology, Japan.


\end{document}